\documentclass[12pt,showpacs,preprintnumbers,amsmath,amssymb]{revtex4}
\usepackage{graphicx}
\usepackage{subfigure}
\begin{document}

\title{Multicriticality in the Blume-Capel model under 
a continuous-field probability distribution}

\vskip \baselineskip

\author{Octavio D. Rodriguez Salmon$^{1}$}
\thanks{E-mail address: octavior@cbpf.br}

\author{Justo Rojas Tapia$^{2}$ }
\thanks{E-mail address: jrojast@unmsm.edu.pe}

\address{
$^{1}$ Centro Brasileiro de Pesquisas F\'{\i}sicas  and National Institute of Science and Technology for Complex Systems,\\
Rua Xavier Sigaud 150 22290-180  Rio de Janeiro,  Brazil \\
$^{2}$ Universidad Nacional Mayor de San Marcos, Facultad de Ciencias F\'{\i}sicas \\
 Av. Venezuela s/n, Apartado Postal 14-0149, Lima-1, Per\'{u}.
}

\date{\today}

\vskip \baselineskip
\vspace{0.4cm}
\begin{abstract}

\noindent

The multicritical behavior of the Blume-Capel model  with infinite-range interactions
is investigated by introducing quenched disorder in the crystal field $\Delta_{i}$, which is 
represented by a superposition of two Gaussian distributions with 
the same width $\sigma$, centered at   $\Delta_{i} = \Delta$ and $\Delta_{i} = 0$,
with probabilities $p$ and $(1-p)$, respectively.  A rich variety of phase
diagrams is presented, and their distinct topologies are shown for
different values of $\sigma$ and $p$. The tricritical behavior is analyzed through the existence of fourth-order critical points, as well
as how the complexity of the phase diagrams is reduced by the strength of the disorder. 

\vspace{0.5cm}


\end{abstract}

\maketitle


\section{Introduction}

The effect of  disorder on  different types of condensed matter orderings is nowadays a subject of considerable interest \cite{binder,ghosal}. For the case of disordered magnetic systems, random-field spin models have been systematically studied, not only for theoretical interests, but for some identifications with experimental realizations \cite{belanger}.  An interesting issue, is the study of how  quenched randomness
destroys some types of criticalities. So, in what concerns the effect produced by random fields in  low dimensions, it has been noticed \cite{hui,berker} that   first-order transitions will be replaced by continuous transitions, so tricritical points and critical end points will be depressed in temperature, and a finite amount of disorder will suppresse them. Nevertheless, in two dimensions, an infinitesimal amount of field randomness seems to destroy  any first-order transition \cite{wehr, boechat}. Interestingly, the simplest model exhibiting a tricritical phase diagram in the absence of randomness is the Blume-Capel model.   
The Blume-Capel model \cite{blume,capel} is a regular Ising model for spin-1  used to  model   $ \bf ^{4}He-^{3}He$ mixtures\cite{emeryg}. The interesting feature is the existence of a  tricritical point in the phase diagram represented in the plane temperature versus crystal field, as shown in Figure 1. This phase diagram was firslty obtained in the mean-field approach, but the same qualitative properties were also observed in low dimensions. The latter was confirmed through some approximation techniques as well as by Monte Carlo simulations \cite{mahan,jain,grollau,kutlu,seferoglu}. Also, the tricritical behavior is still held  in two dimensions \cite{clusel,care,paul,caparica}. Nevertheless, in other models this situation is controversial. For example,  the random-field Ising Model in the mean-field approach \cite{aharony} also exhibits a tricritical point, but  some Monte Carlo simulations \cite{fytas} in the cubic lattice suggest that this is only an artifact of the mean-field calculations. Accordingly, this interesting fact in the Blume-Capel model  motivated some authors to explore the richness of this model, within the mean-field approach,  by introducing disorder  in the crystal field \cite{hamid,benyoussef,salinas,carneiro} as well as by adding and external random field \cite{miron}. For the former case, it was obtained  a variety of phase diagrams including different critical points  with some similar topologies found for the random-field spin$-1/2$ Ising model \cite{kaufman,octavio}.  However, in those studies the fourth-order critical points, which limit the  existence of tricritical points, were overlooked. Consequently, our aim  in this work is to   improve those previous studies by  considering a more general probability distribution function for the crystal field,  and bettering some results given in references \cite{benyoussef,salinas,carneiro}. The next section is dedicated to define the model and the special critical points produced by it.
\vskip \baselineskip

\begin{figure}[htp]
\begin{center}
\includegraphics[height=5.5cm]{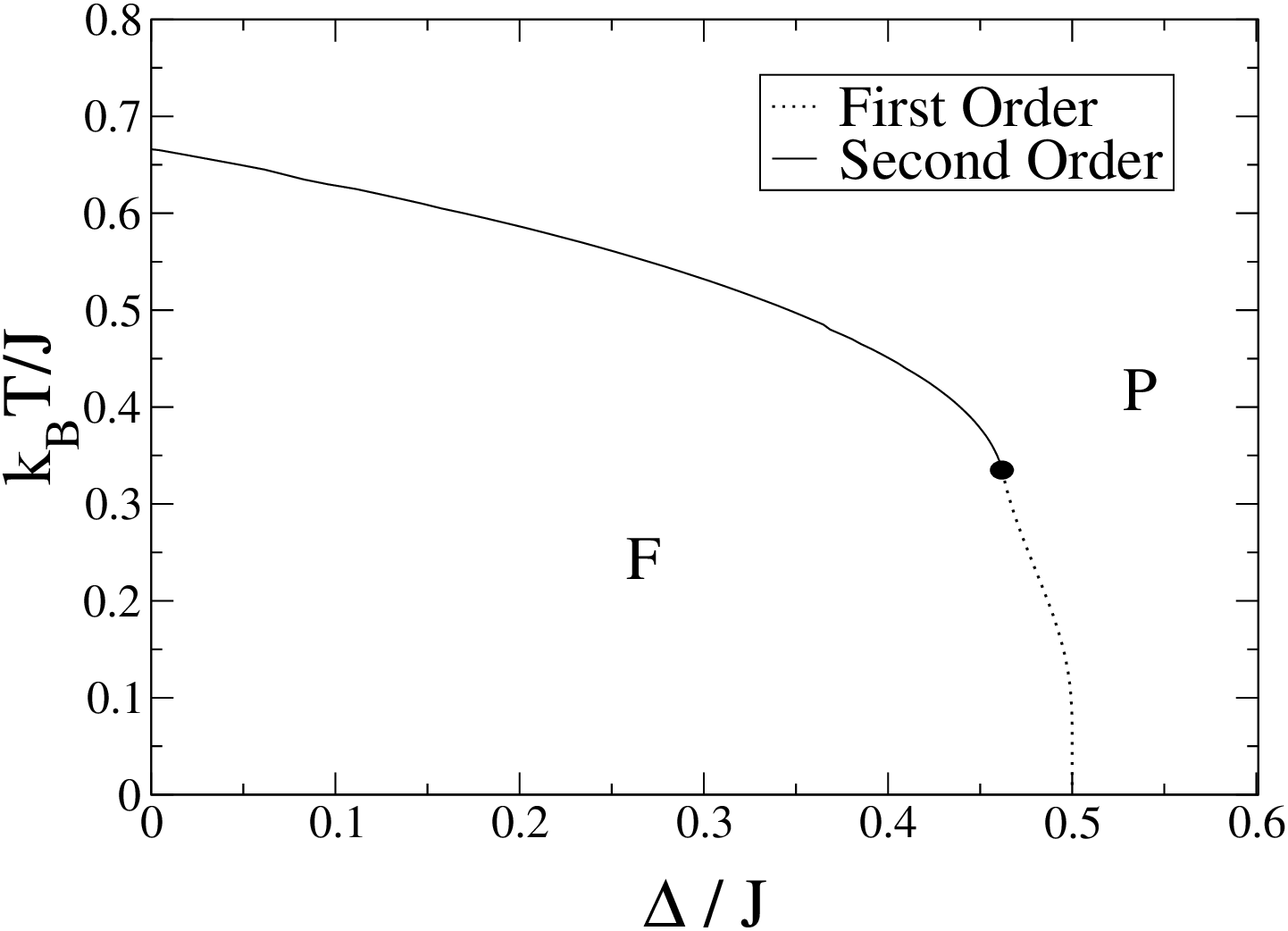}
\end{center}
\caption[]{\footnotesize Phase diagram of the Blume-Capel model in the plane $k_{B}T/J-\Delta/J$ within the mean-field approach, where $k_{B}$ is the Boltzman constant, $T$ is the temperature, $J>0$ is the coupling constant between each pair of spins, and $\Delta$ is the crystal field (also called the anisotropy field). The black circle represents the tricritical point. The Ferromagnetic and Paramagnetic phases are represented by $\bf F$ and $\bf P$, respectively. The full line represents the continuous or second-order critical frontier, and the dotted
line is for the first-order frontier.}
\label{n}
\end{figure}  
\vskip \baselineskip
   
\section{The Model}   
   
The infinite-range-interaction Blume-Capel model is given by the following
Hamiltonian

\begin{equation}  {\cal H} = -\frac{J}{N}\sum_{(i,j)} S_{i}S_{j} + \sum_{i} \Delta_{i} S_{i}^{2}  ~, \label{Ham} \end{equation}
\vskip \baselineskip
where $S_{i}=-1,0,1$, and $N$ is the number of spins. The first sum runs over all distinct pairs of spins.
The coupling constant $J$ is divided by $N$ in order to maintain the extensivity.
The crystal fields are represented  by  quenched variables $\{ \Delta_{i} \}$,  obeying
the  probability distribution function (PDF) given by, 

\begin{equation}P(\Delta_{i}) =    \frac{p}{\sqrt{2\pi} \,\sigma} \exp \left[ - \frac{(\Delta_{i}-\Delta)^{2}}{2{\sigma}^{2}} \right] + \frac{(1-p)}{\sqrt{2\pi} \,\sigma} \exp \left[ - \frac{{\Delta_{i}}^{2}}{2{\sigma}^{2}} \right ] ~, \label{PDF} \end{equation}
\vskip \baselineskip
which consists of a superposition of two independent Gaussian distributions
with the same width $\sigma$, centered at  $\Delta_{i}=\Delta$ and $\Delta_{i}=0$, with probabilities $p$ and $(1-p)$, respectively.
For $\sigma =0$, we recover the bimodal distribution studied in references \cite{benyoussef,salinas}, and
for  $p=1$, the simple Gaussian one of reference \cite{carneiro} . For $\sigma=0$ and $p=1$, we go back to the simple Blume-Capel model without randomness\cite{emeryg}. \\
By standard procedures \cite{octavio}, we   get the analytical expression for the free energy per spin ($f$), through which may be obtained a self-consistent equation for the magnetization $m$. Thus, we have the  following relations at the equilibrium,
\begin{equation} f = \frac{1}{2}Jm^{2} 
-\frac{1}{\beta} E \left \{ \log(2\exp(-\beta\Delta_{i}) \cosh(\beta J m) +1) \right \} ~, \label{f} \end{equation}  

\begin{equation} m = \sinh(\beta m) E \left \{ {\left [  \cosh(\beta m) +\frac{1}{2} \exp(\beta \Delta_{i})   \right ]}^{-1} \right \} ~, \label{m} \end{equation}
\vskip \baselineskip
where the quenched average, represented by $E \{...\}$, is taken with respect to the PDF given
in Eq. (\ref{PDF}), and $\beta = 1/(k_{B}T)$. To write conditions for locating  tricritical and fourth-order critical points, we expand the right hand of Eq. (\ref{m}) in powers of $m$ (Landau's expansion, see \cite{stanley}). Conveniently, we expand the magnetization up to seventh order in $m$, so

\begin{equation} m= A_{1} m + A_{3}m^{3}+A_{5}m^{5}+A_{7}m^{7}+...  ~,\end{equation}

where

\begin{equation} A_{1} =  \beta E\{ g_{i} \} ~,\end{equation}
\begin{equation} A_{3} = \beta^{3} E\{(\frac{1}{6}g_{i} -\frac{1}{2}g_{i}^{2}) \} ~, \end{equation}
\begin{equation} A_{5} = \beta^{5} E\{  (\frac{1}{120}g_{i} - \frac{1}{8}g_{i}^{2}  +\frac{1}{4}g_{i}^{3}) \} ~, \end{equation}

\begin{equation}   A_{7} = \beta^{7}  E\{  ( \frac{1}{5040}g_{i} -\frac{1}{80}g_{i}^{2} + \frac{1}{12}g_{i}^{3} -\frac{1}{8}g_{i}^{4} ) \} ~,\end{equation}
and
\begin{equation}  g_{i} =  (1+\frac{1}{2} \exp(\beta \Delta_{i}))^{-1} ~. \end{equation}
\vskip \baselineskip

In order to obtain the continuous critical frontier one sets $A_{1}=1$, provided
that $A_{3} <0$. If a first-order critical frontier begins after the continuous one, the latter
line ends at a tricritical point if $A_{3}=0$, provided that $A_{5}<0$. The possibility of a fourth-order critical point is given for $A_{1}=1$, $A_{3}=0$,  $A_{5}=0$ and $A_{7}<0$. Thus, a fourth-order point may be regarded as the last
 tricritical point. \\
By taking  $\beta \to \infty$ ($T \to 0$), we get the asymptotic limit of Eqs. (\ref{f}) and (\ref{m}), so we have

\begin{eqnarray}\nonumber
 f & = & \frac{1}{2} Jm^{2}  -   p\left(  \frac{1}{2} (Jm-\Delta) \left ( 1+ {\rm erf} \left [ \frac{Jm-\Delta}{\sqrt{2} \, \sigma} \right ] \right ) +\frac{\sigma}{\sqrt{2\pi}} \exp  \left [ - \frac{(Jm-\Delta)^{2}}{2 \, {\sigma}^{2}}   \right ]  \right ) \nonumber \\
 & - &  (1-p) \left (  \frac{1}{2} Jm \left ( 1+ {\rm erf} \left[\frac{Jm}{\sqrt{2} \, \sigma} \right ] \right) +\frac{\sigma}{\sqrt{2\pi}} \exp  \left [ - \frac{J^{2} m^{2}}{2 \, {\sigma}^{2}}   \right ]  \right )
~, \end{eqnarray}

\begin{equation}  m = \frac{p}{2} \left ( 1+ {\rm erf} \left [  \frac{Jm -\Delta}{\sqrt{2}\, \sigma} \right ]  \right ) + \frac{(1-p)}{2} \left ( 1+ {\rm erf} \left [  \frac{Jm}{\sqrt{2}\, \sigma} \right ] \right )  ~,
  \end{equation}

where 

\begin{equation} {\rm erf } \left ( \frac{x}{\sqrt{2}} \right  ) = \sqrt{\frac{2}{\pi}} \int_{0}^{x} dz e^{-z^{2}/2} ~. \end{equation}
\vskip \baselineskip
The critical frontiers, for a given pair $(\sigma,p)$,  are obtained by  solving a non-linear set of equations, which consist of equating the free energies for the corresponding phases (Maxwell's construction), and the respective magnetization equations based on the relations given in Eqs. (\ref{f}), and (\ref{m}). We must carefully verify that every numerical solution  minimizes the free energy. \\
The symbols used to represent the different critical  lines and points \cite{octavio} are as follows:

\begin{itemize}

\item Continuous or second-order critical frontier: continuous line;

\item Fist-order critical frontier: dotted line;

\item Tricritical point: located by a black circle;

\item Fourth-order critical point: located by an empty square;

\item Ordered critical point: located by an asterisk;

\item Critical end point: located by a black triangle.

\end{itemize} 

To clarify, we mean by a continuous critical frontier that which separates two distinct phases
through which the order parameter changes continuously to pass from one phase to another, contrary to the case of the first-order transition,
through which, the order parameter suffers a discontinuous change, so the  two corresponding phases coexist at  each critical point.
A tricritical point is basically the point in which a continuous line terminates to give rise a first-order critical line.
A fourth-order critical point is sometimes called a vestigial tricritical point, because it may be regarded as the last tricritical point. An ordered critical point is the point, inside an ordered region, where  a first-order critical line  ends, above which  the order parameter passes smoothly from one ordered phase to  the other.
Finally, a critical end point corresponds to the intersection of a continuous
line that separates the paramagnetic from one of the ferromagnetic phases with a first-
order line separating the paramagnetic and the other ferromagnetic phase. In following section we make use of this
definitions. 

\section{Results and Discussion}

The distinct  phase diagrams  for the present model were numerically obtained by scanning the whole p-domain for each $\sigma$-width. So, distinct topologies  belonging to different $p$-ranges were found for a given $\sigma$.  For instance, Figure 2 shows the whole  variety of them  for a small $\sigma/J=0.1$, for each arbitrary  representative  $p$.

\begin{figure}[htp]

\centering
\subfigure[][] {\includegraphics[width=5.0cm]{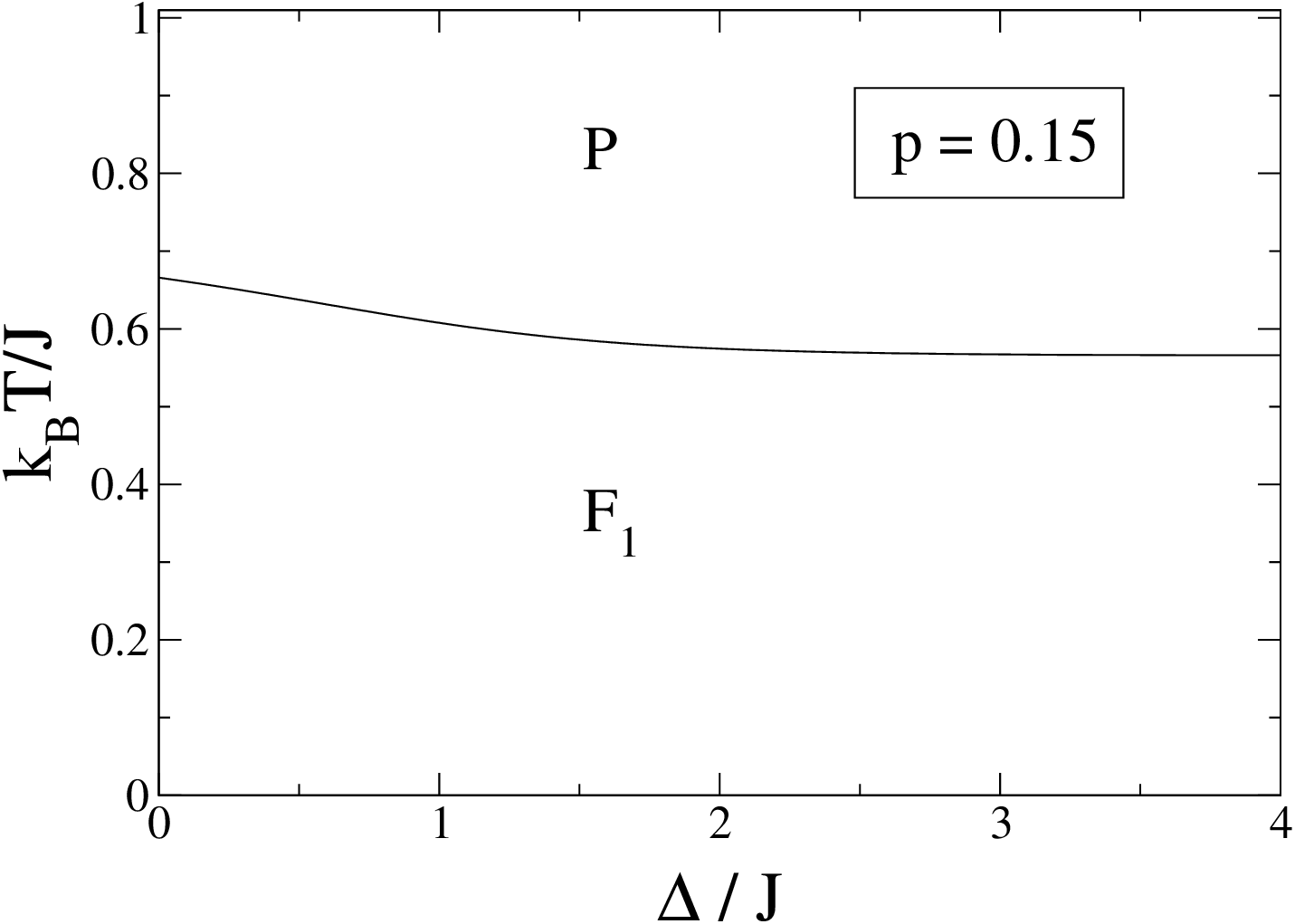}}
\vspace{0.7cm} 
 \subfigure[][] {\includegraphics[width=5.0cm]{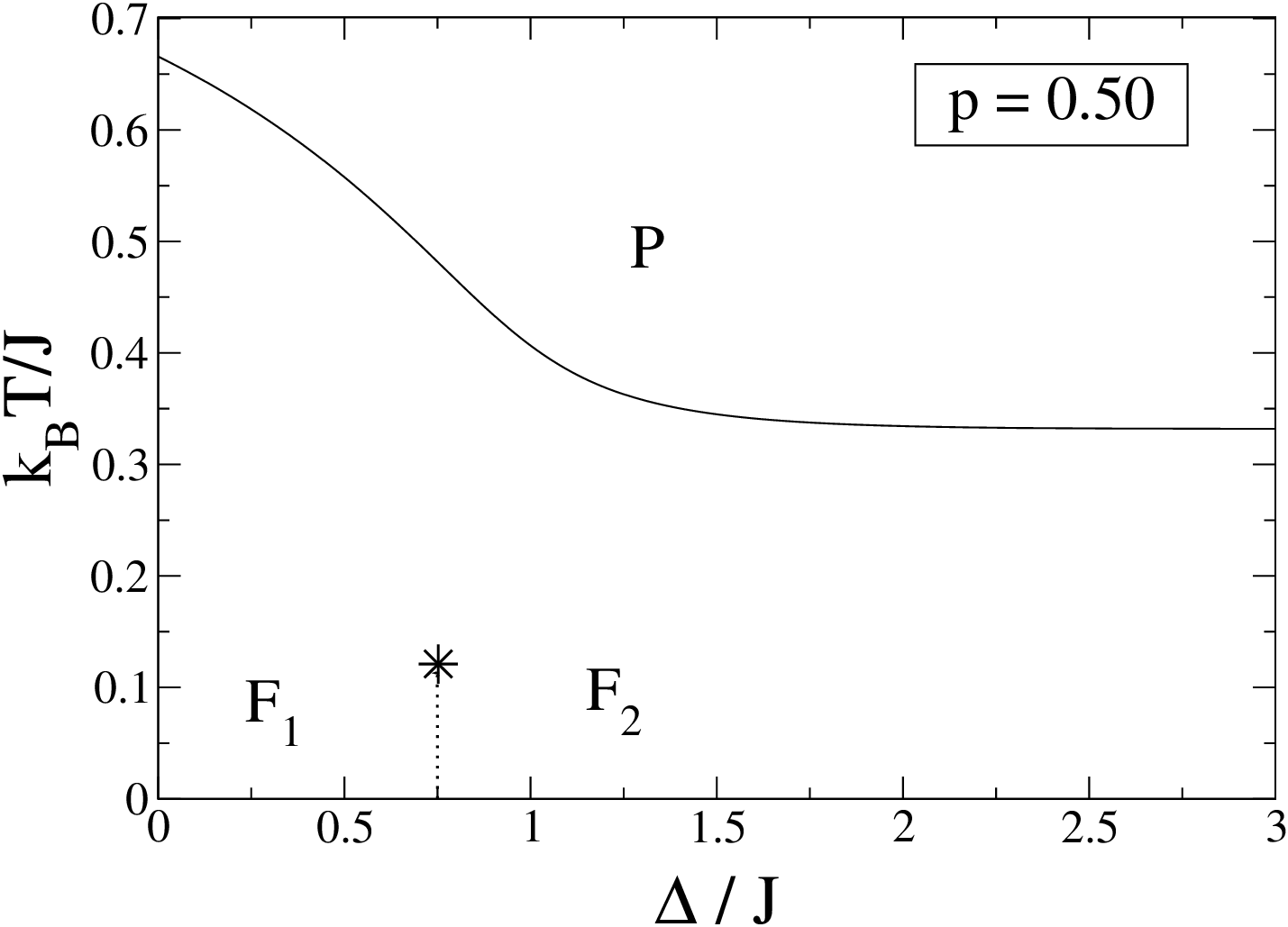}}
\vspace{0.4cm} 
 \subfigure[][]{\includegraphics[width=5.0cm]{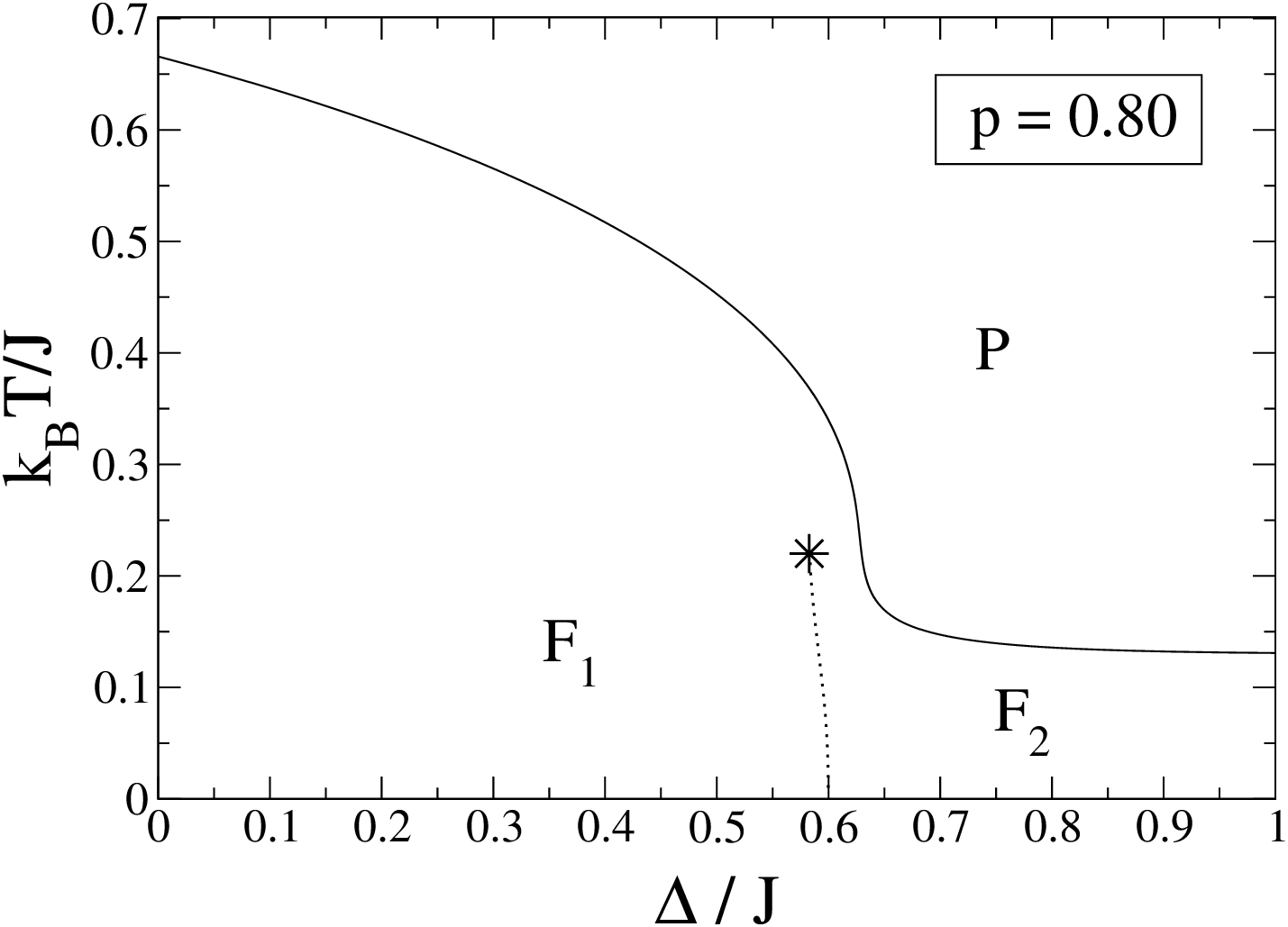}}
\vspace{0.4cm}   
\subfigure[][] {\includegraphics[width=5.0cm]{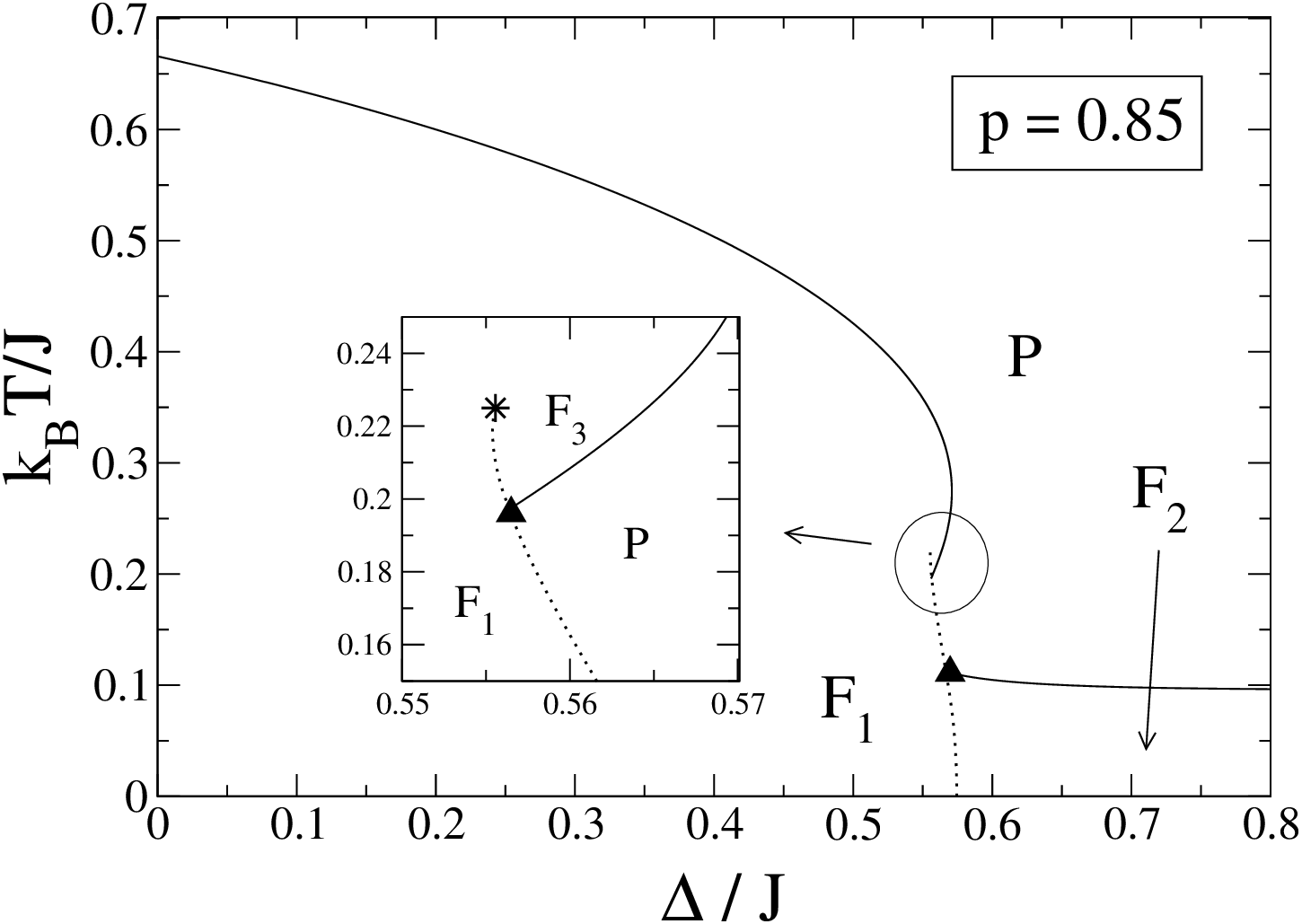}}
 \subfigure[][]{\includegraphics[width=5.0cm]{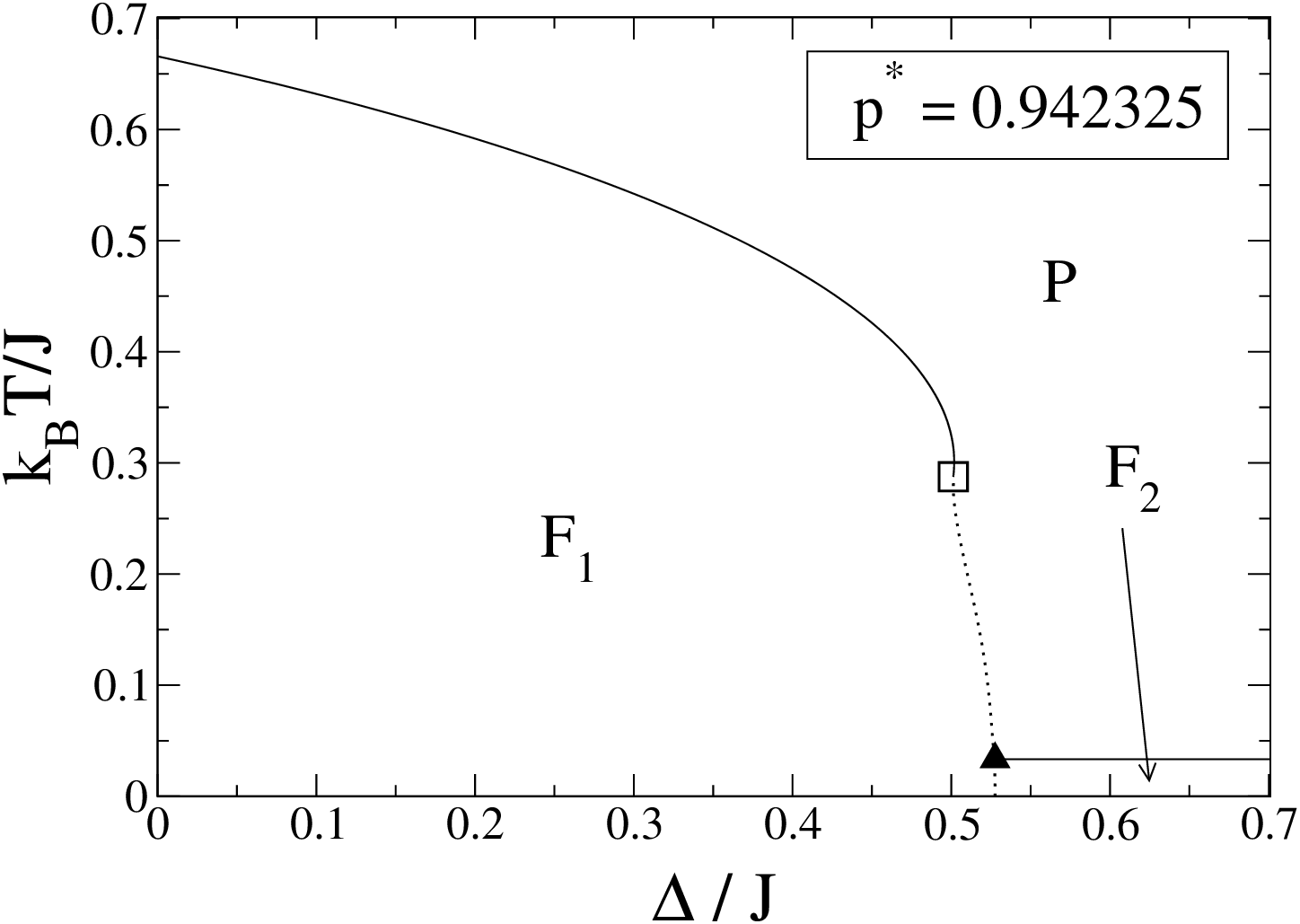}}
\subfigure[][]{\includegraphics[width=5.0cm]{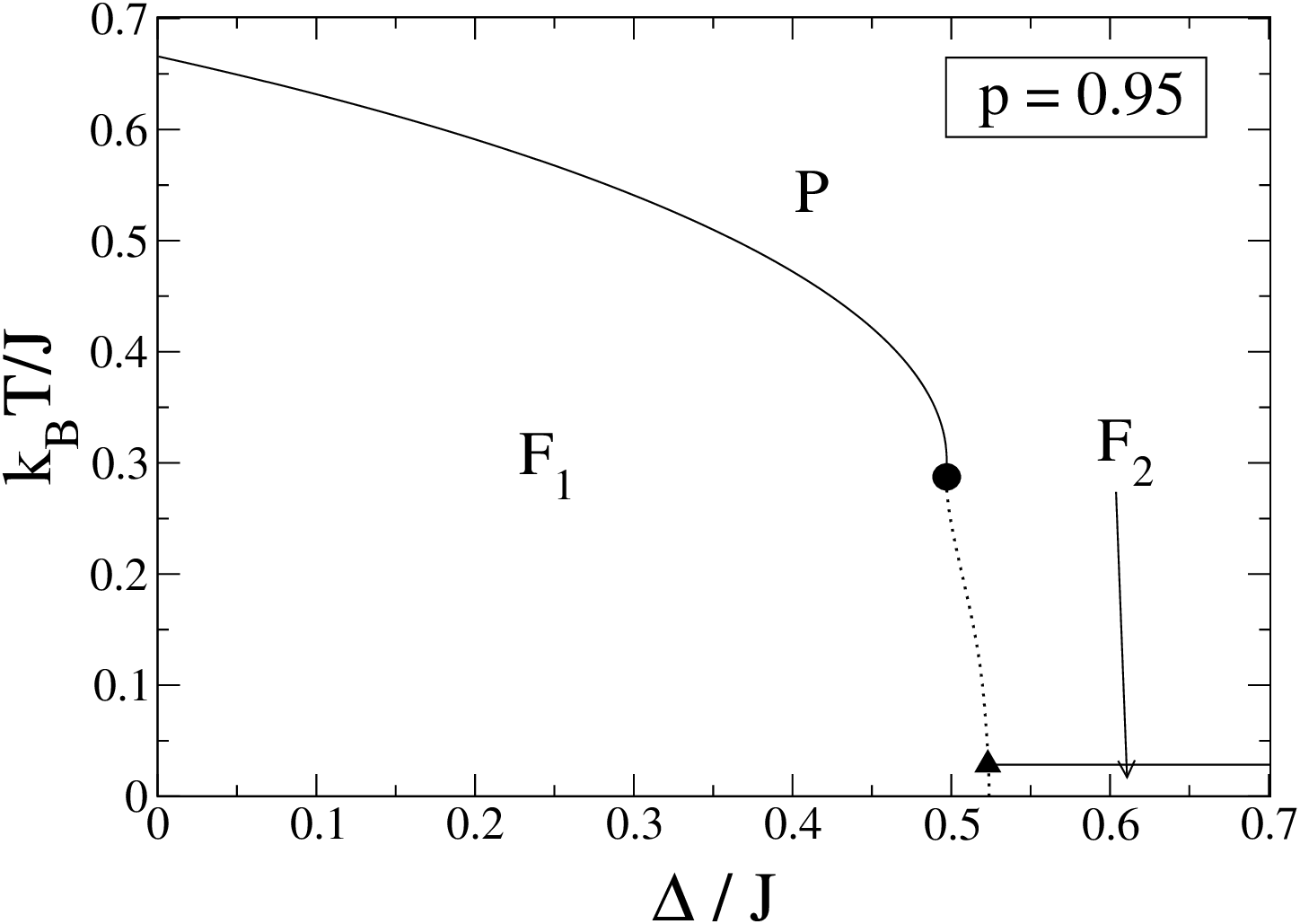}}
 
\caption{\footnotesize Phase diagrams of the Blume-Capel Model whose crystal field obeys the PDF given in Eq.(\ref{PDF}). For $\sigma/J=0.1$, the diagrams show a variety of topologies according to the probability $p$. For convenience, we classify them in four topologies, so in (a) is shown Topology I; in (b) and (c) Topology II; in (d) Topology III, then figures (e) and (f) represent Topology IV.  }

\label{imagenes}
\end{figure}

Note that for  small values of $p$,  one only ferromagnetic order appears at low temperatures, as shown in Figure 2(a) for $p=0.15$. We designate it as  Topology I.  Figures 2(b) and (c)  ($p=0.5,0.8$)  represent the same  topology (Topology II), which consists of one first-order critical line separating two different ferromagnetic phases $\bf F_{1}$ and $\bf F_{2}$, and a continuous line remaining for $\Delta/J \to \infty$. Figure 2(c), though qualitatively the same as in 2(b), is intended to show how the first-order line and the continuous line approach themselves as $p$ increases.  Figure 2(d) shows Topology III, for $p=0.85$, so the preceding first-order line is now dividing the continuous line by two critical end points. Note that the upper continuous line  terminates, following a reentrant path, at a critical end point where the phases  $\bf F_{1}$,  $\bf F_{3}$, and $\bf P$ coexist. So, at the lower critical end point, $\bf F_{1}$,  $\bf F_{2}$, and $\bf P$ coexist. Above the ordered critical point, the order parameter  passes smoothly from $\bf F_{1}$ to $\bf F_{3}$ (see the inset there). If we increase $p$ 
up to some $p=p^{*}$, the upper continuous line and the first-order line will be met by a fourth-order point (represented by a square) as shown in Figure 2(e). Thus, $p^{*}$ is the threshold for Topology IV. Then, for  $p > p^{*}$, those lines will be met at a tricritical point, as noticed in Figure 2(f). Conversely,  tricritical points  appear  for $p>p^{*}$, so the last one  for $p=p^{*}$. The same types of phase diagrams are found in references \cite{benyoussef,salinas,carneiro}. Nevertheless, we improve their results, not only bettering some of their numerical calculations, but  in that we may now locate the regions of validity of these topologies in the plane $\sigma/J-p$. To this end, we start  by locating  the fourth-order points in the plane $\sigma/J-p$, as shown in Figure 3.

\begin{figure}[htp]
\begin{center}
\vspace{0.7cm} 
\includegraphics[height=6cm]{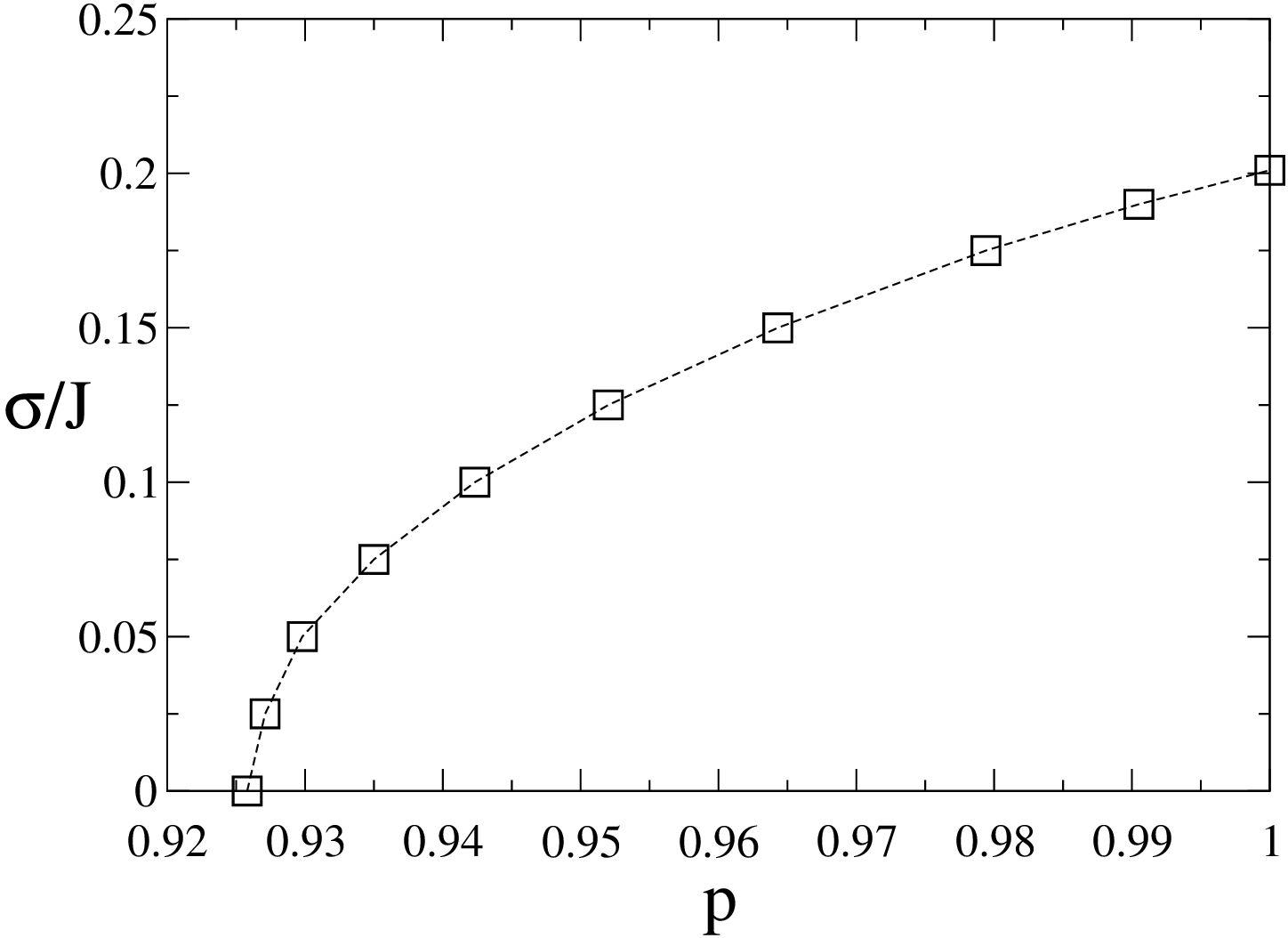}
\end{center}
\caption[]{\footnotesize Some fourth-order critical points located in the plane $p-\sigma/J$. Note that if $\sigma = 0$, we recover the bimodal case studied in references \cite{benyoussef, salinas}, where we found $p^{*}=0.9258$, just in agreement with them. Note that if $p=1$,  $\sigma/J = 0.202$, so it is a $\sigma$-limit for the tricritical behavior. The dashed line is only a guide to the eyes.}
\label{nua}
\end{figure}  


Note that   $\sigma/J=0.202$ is a cut-off for the tricritical behavior. Then, Topology IV will no longer found for greater widths. On the other hand, we determine the threshold for Topology III, by  estimating numerically which value of $p$, for each $\sigma/J$, produces a situation like that presented in Figure 4 (case $\sigma/J=0.1$),  where we see how  Topology III emerges for a $p$ slightly greater than $0.836$. For $\sigma/J=0$, we found this threshold for $p=0.8245$, which is smaller than that obtained in reference \cite{benyoussef}. There, the authors suggested that Topology III disappears for $p<8/9=0.888...$ . However,  Figure 5 illustrates that this type of phase diagram 
is still present even for a smaller $p$, as confirmed by the free energy evaluated at  three disctinct ($k_{B}T/J,\Delta/J$)-points along the first-order critical line, at which there are three types of coexistences, namely,  $\bf F_{1}$ with $\bf F_{3}$, $\bf F_{1}$ with $\bf P$, and $\bf F_{1}$ with $\bf F_{2}$. We also noted another discrepancy  with respect to a critical $\sigma/J$, found in reference \cite{carneiro}, above which Topology III disappears for $p=1$. There, the authors affirmed that if  $\sigma/J > 0.229$, the paramagnetic-ferromagnetic transition becomes second order at all temperatures, but we noticed that it only happens for a greater width, namely, $\sigma/J=0.283$. \\
In order to obtain the  frontier  which separates Topologies I and II (in the plane $\sigma/J-p$), we have to find the corresponding $p$, for a given $\sigma/J$, that locates the one ordered critical point  at $T=0$. To this end, the next subsection is focused on zero temperature calculations.  
 
\vskip \baselineskip  
 \begin{figure}[htp]
\centering
\vspace{0.7cm} 
\subfigure[][] {\includegraphics[width=6.0cm]{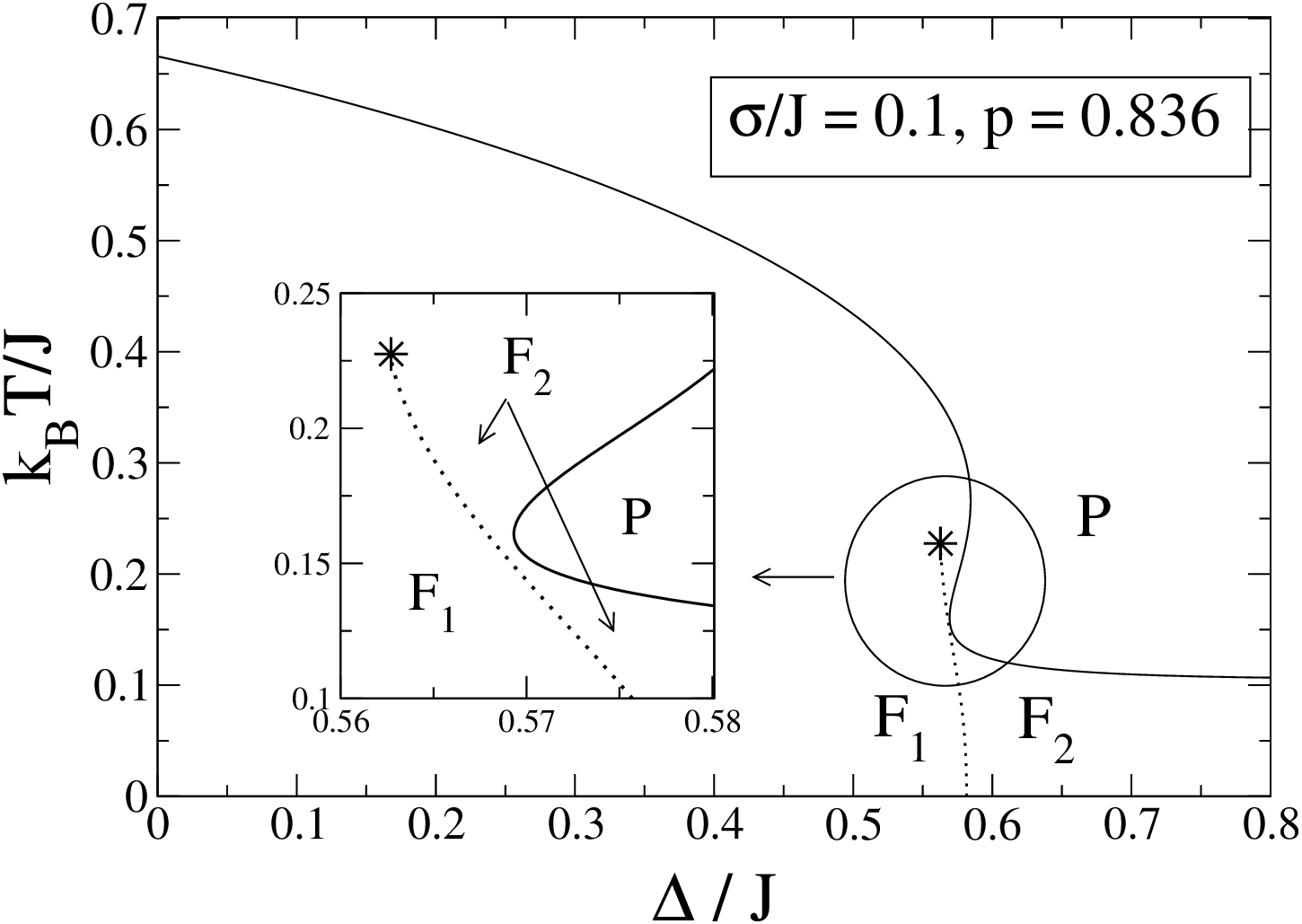}}
 \subfigure[][] {\includegraphics[width=6.0cm]{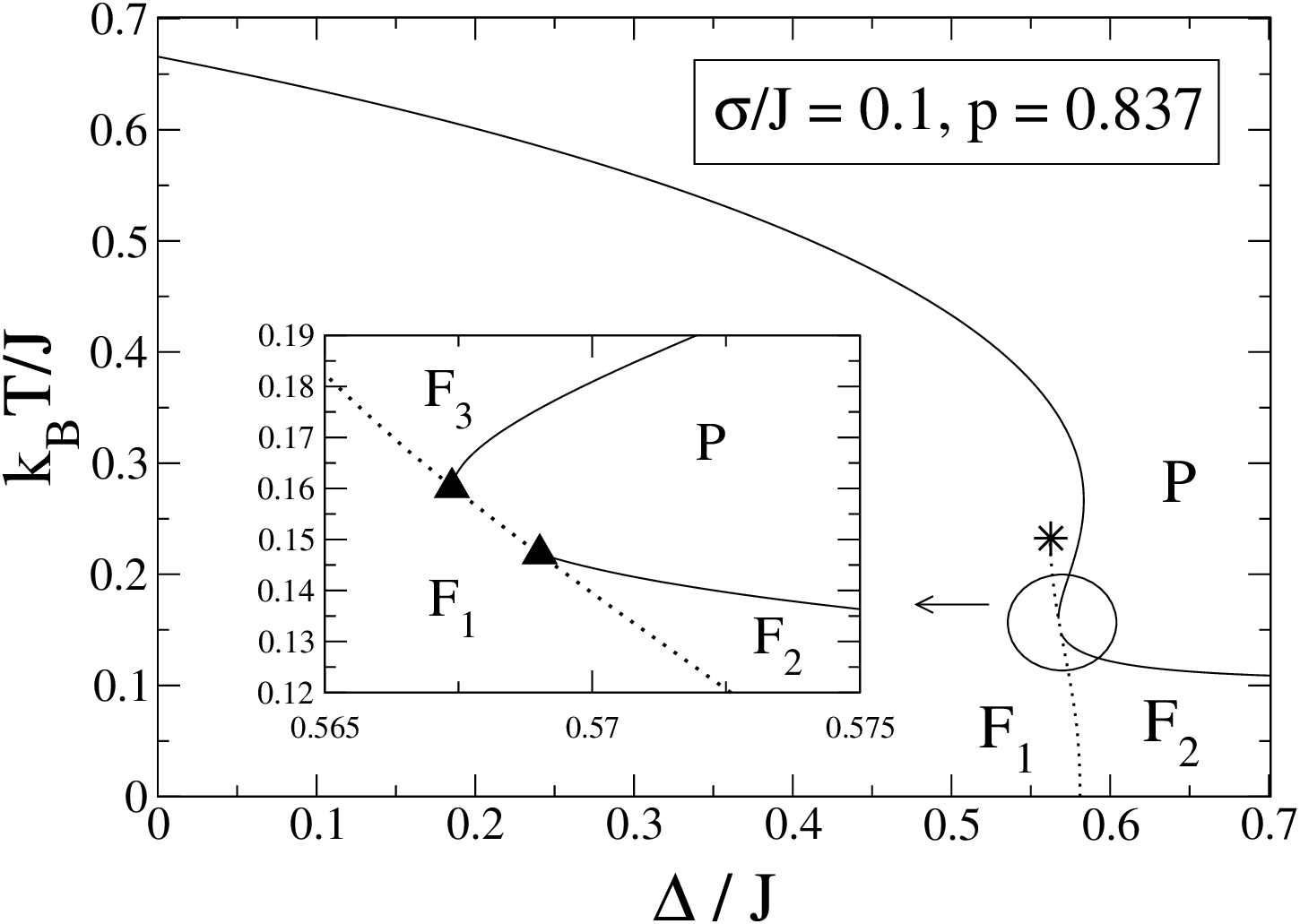}}
\caption{\footnotesize Phase diagrams (for $\sigma/J = 0.1$) showing two slightly different values of $p$, between which there is
a critical $p$ for passing from Topology II to III. So, that critical point must be found for $p=0.8365 \pm 0.0005$. }
\label{imagenes4}
\end{figure}   
\vskip \baselineskip

\begin{figure}[htp]

\centering
\subfigure[][] {\includegraphics[width=5.0cm]{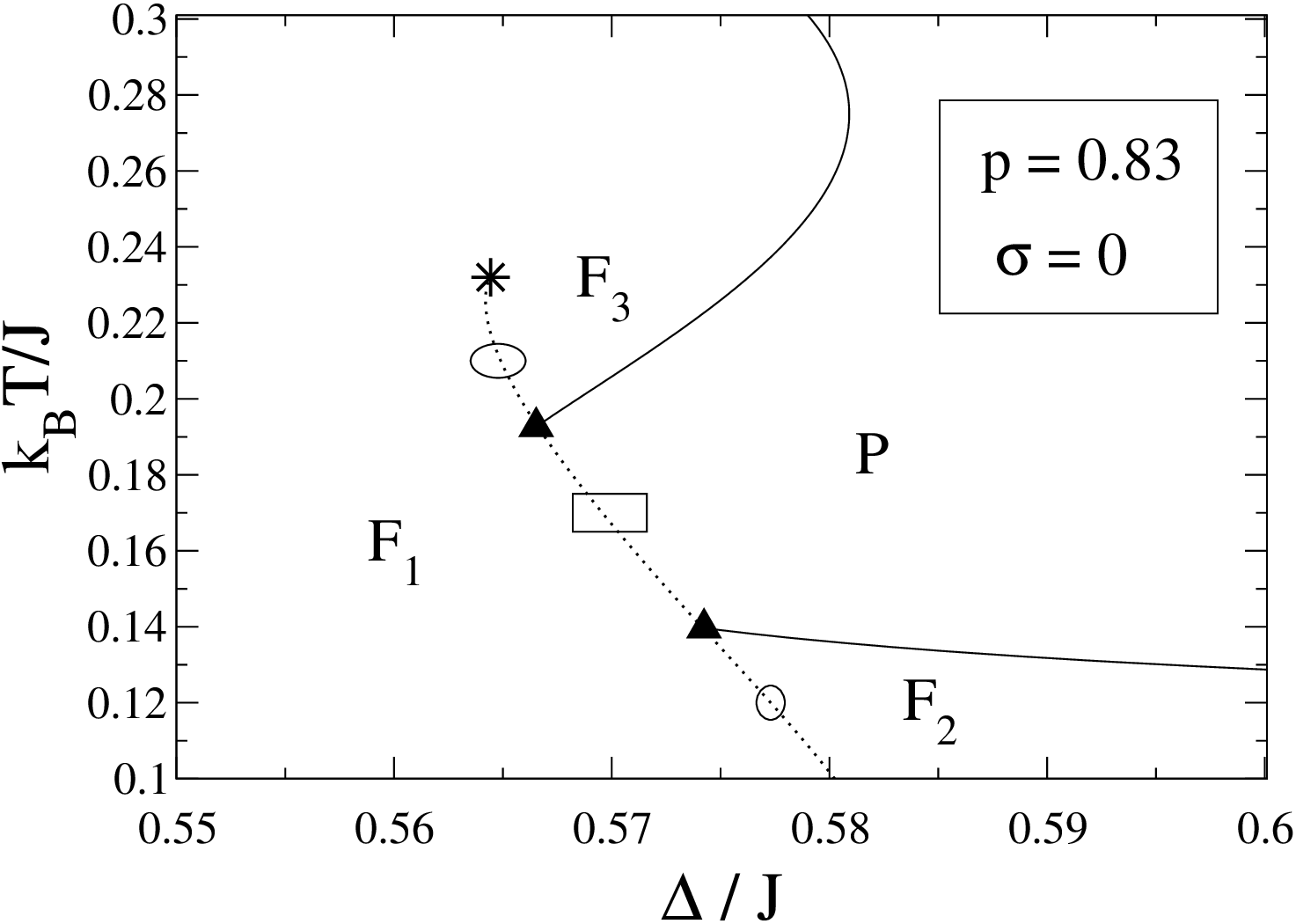}}
\vspace{0.7cm} 
 \subfigure[][] {\includegraphics[width=5.0cm]{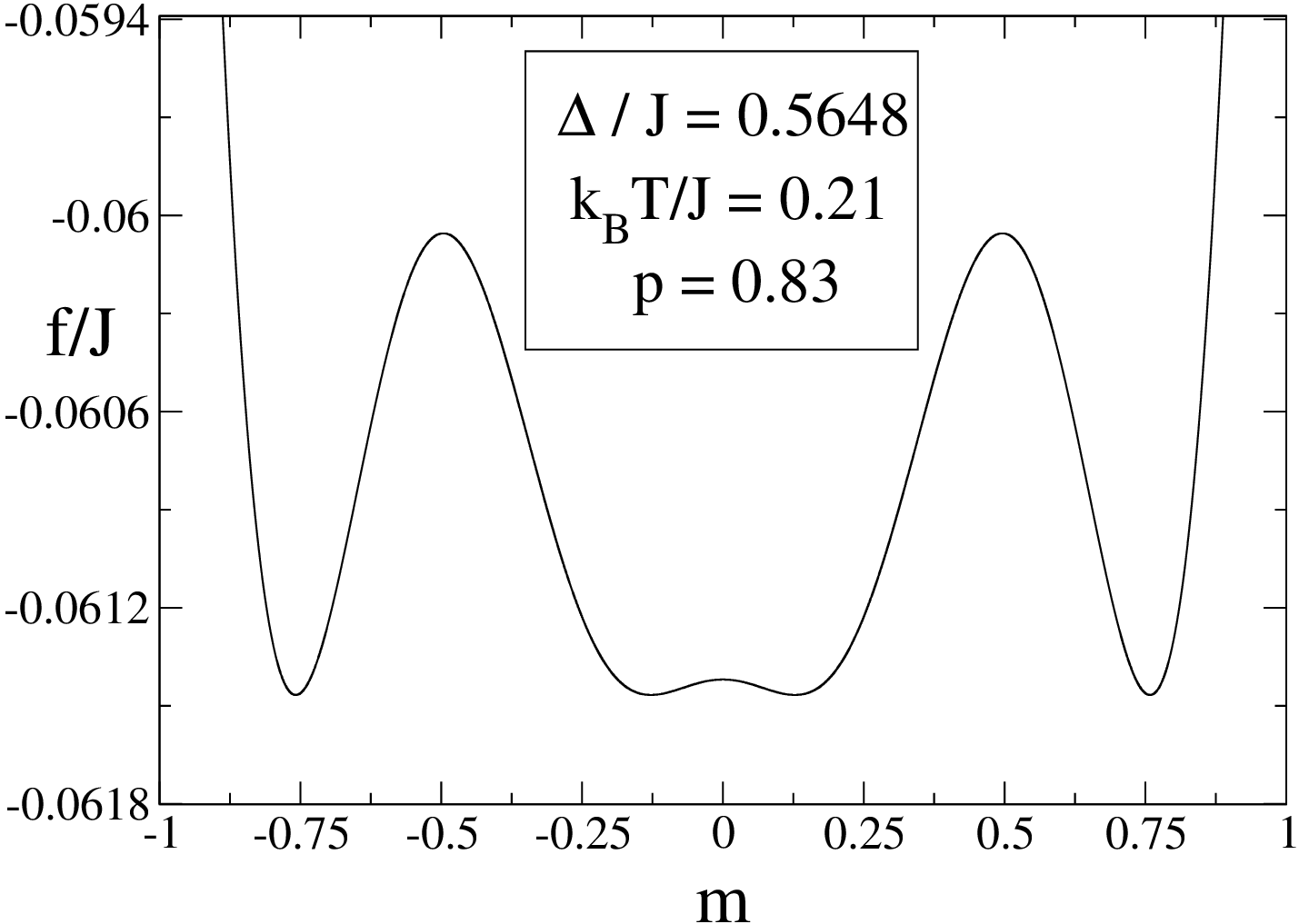}}

\vspace{0.4cm} 
 \subfigure[][]{\includegraphics[width=5.0cm]{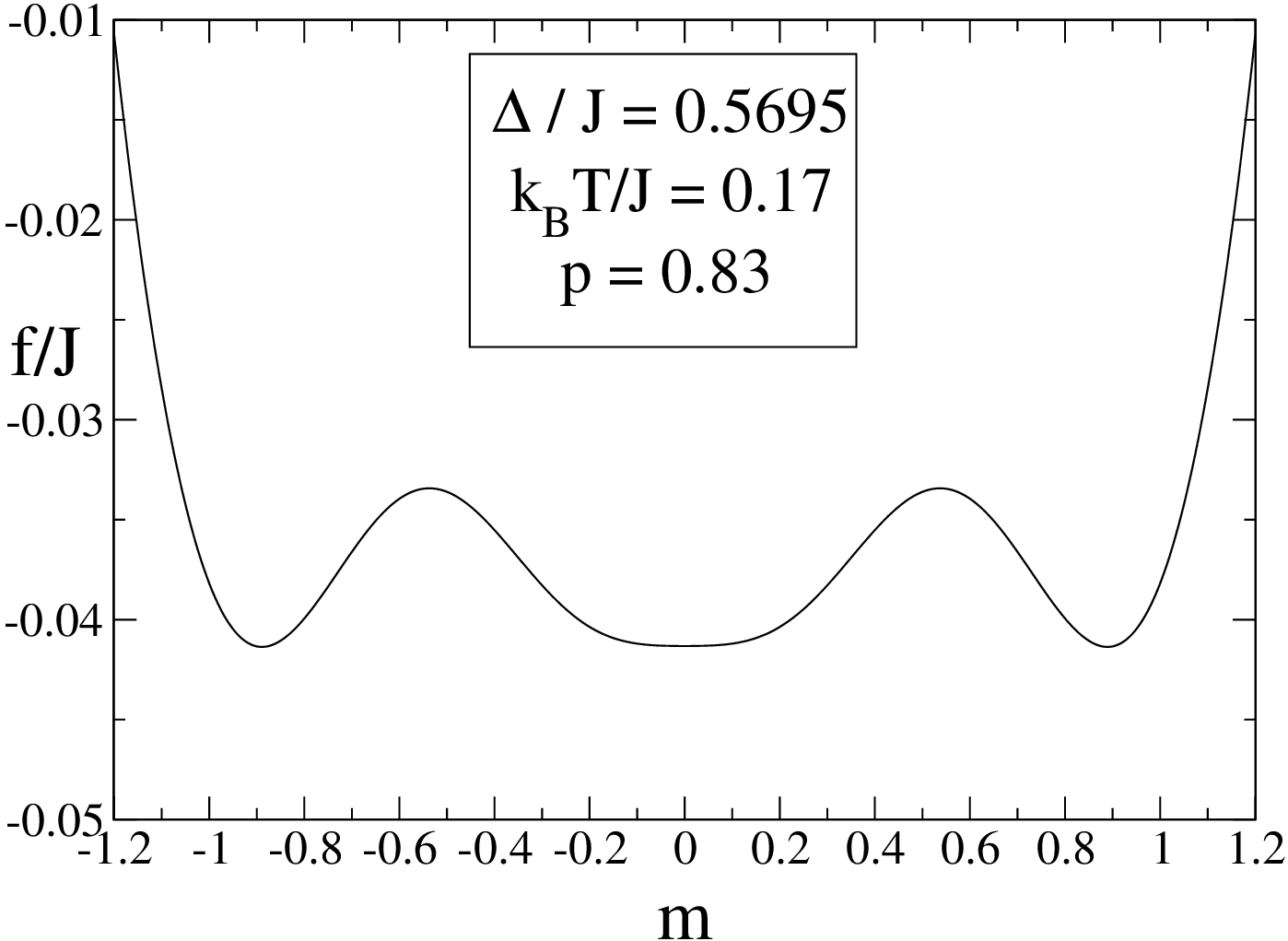}}
\vspace{0.4cm}   
\subfigure[][] {\includegraphics[width=5.0cm]{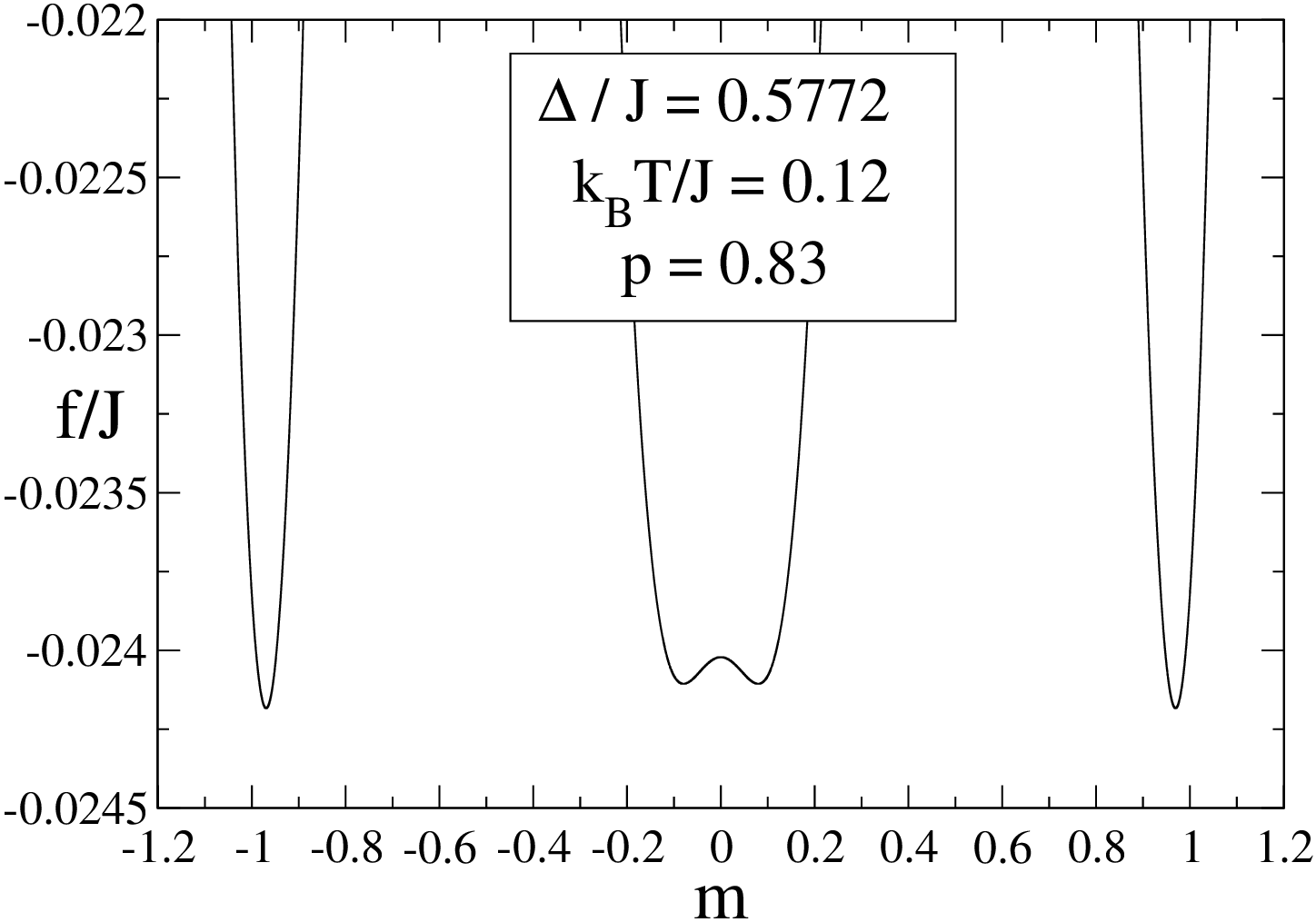}}

\caption{\footnotesize In (a) is shown the most critical region of the phase diagram for $\sigma=0$, and $p=0.83$. It is a typical phase  diagram for Topology III (like that of Figure 2(d)). Note that three points belonging to the first-order critical line are highlighted by an ellipse, a rectangle, and a circle. The ellipse is surrounding a critical point where the phases $\bf F_{1}$ and $\bf F_{3}$ coexist, as confirmed through the free energy versus the magnetization in (b). In (c) and (d) the free energy shows which phases are coexisting at the points surrounded by the rectangle and the circle. Thus, in (c), the phases $\bf F_{1}$ and $\bf P$ coexist, because  two symmetric minima  at  finites values of $m$, and one minimum at $m=0$, are at the same level. In (d), like in (b), are shown four symmetric minima at the same level. Therefore, the phases $\bf F_{1}$ and $\bf F_{2}$ coexist at this critical point.   }

\label{imagenes}
\end{figure}


\subsection{Analysis at $T=0$}

In order to perform zero temperature calculations we make use of the equations (11) and (12). Consequently, for  
$\sigma/J = 0$ (see reference \cite{salinas}), there are two ferromagnetic phases $\bf F_{1}$ and $\bf F_{2}$ coexisting at $\Delta/J = 1-p/2$, having magnetizations $m_{1}=1$ and $m_{2}=1-p$, respectively.  We observed that these relations still remain up to some finite $\sigma$, after which a $\sigma$-dependency emerges. So, for a greater width called $\sigma^{'}$,
the ordered critical point (that of Topology III) must be found at $T=0$. Then, the first-order critical line is supressed and one only  ferromagnetic order exists for any $p$. For instance, if we choose $p=0.5$,  we find  $\sigma^{'}/J=0.2$, as illustrated in Figure 6. There, the zero temperature free energy  versus the order parameter is plotted for three different values of $\sigma/J$,  at the point where $ \bf F_{1}$ and $ \bf F_{2}$ coexist.  Thus, In (a),  two minima are at the same level for $\sigma=0.1$. In (b), it still happens for $\sigma=0.15$. Nonetheless, in (c), for $\sigma/J=0.2$, the ordered critical point is already at $T=0$. Therefore, for this particular $p$, there is only one ferromagnetic phase for $\sigma/J>0.2$. 
 \begin{figure}[htp]
\centering
\vspace{0.7cm} 
\subfigure[][] {\includegraphics[width=5.0cm]{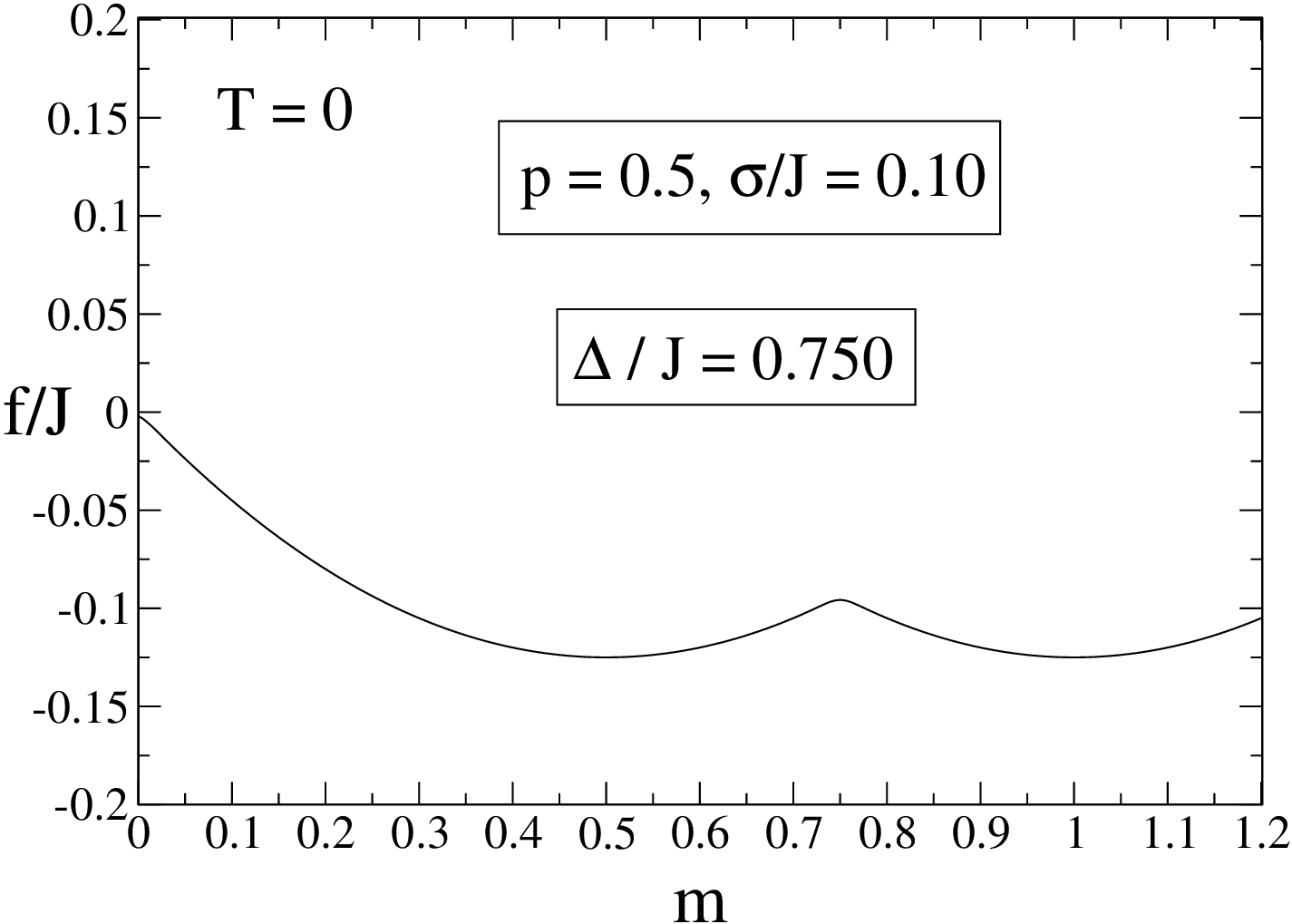}}
 \subfigure[][] {\includegraphics[width=5.0cm]{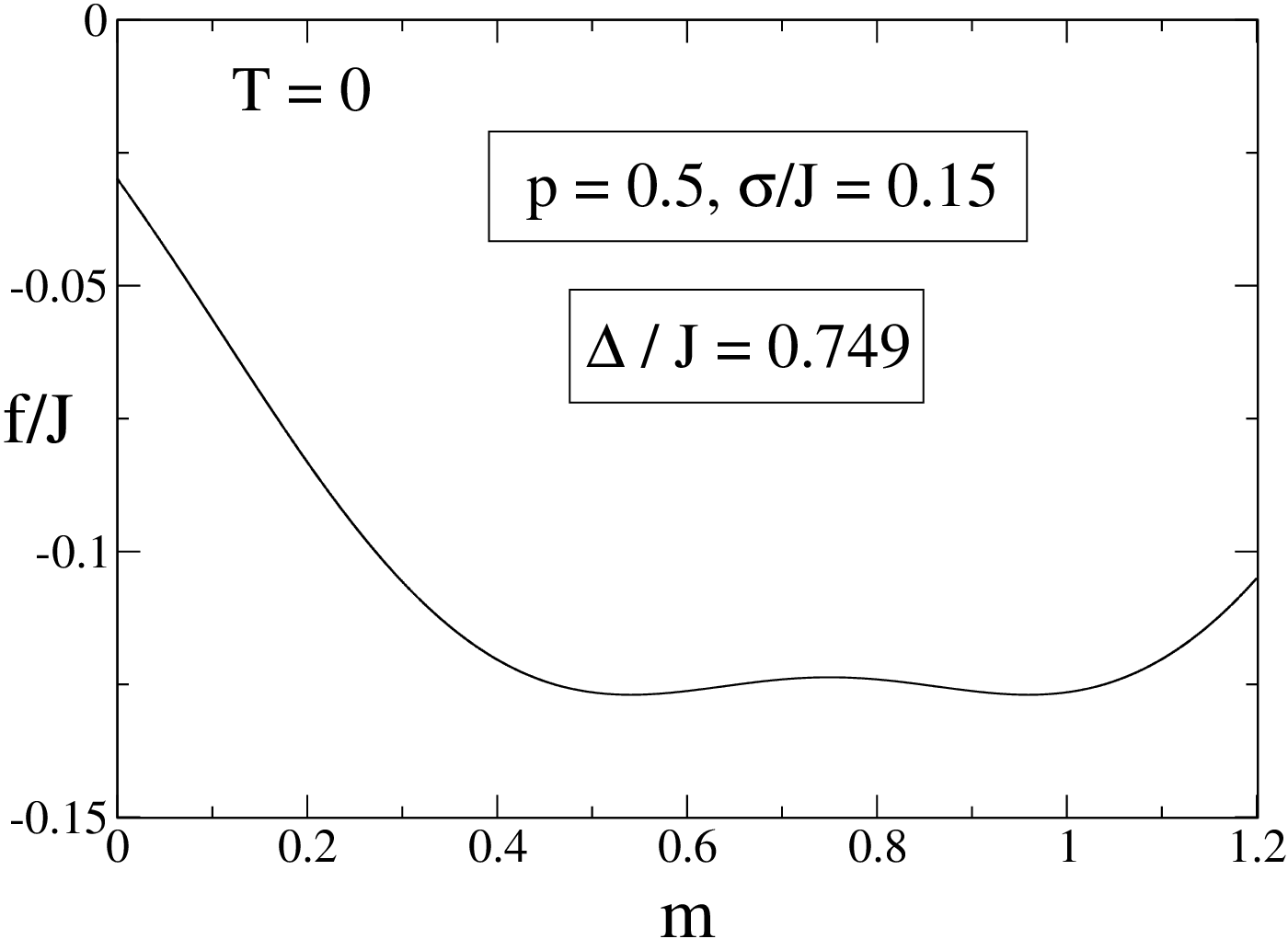}}
 \subfigure[][]{\includegraphics[width=5.0cm]{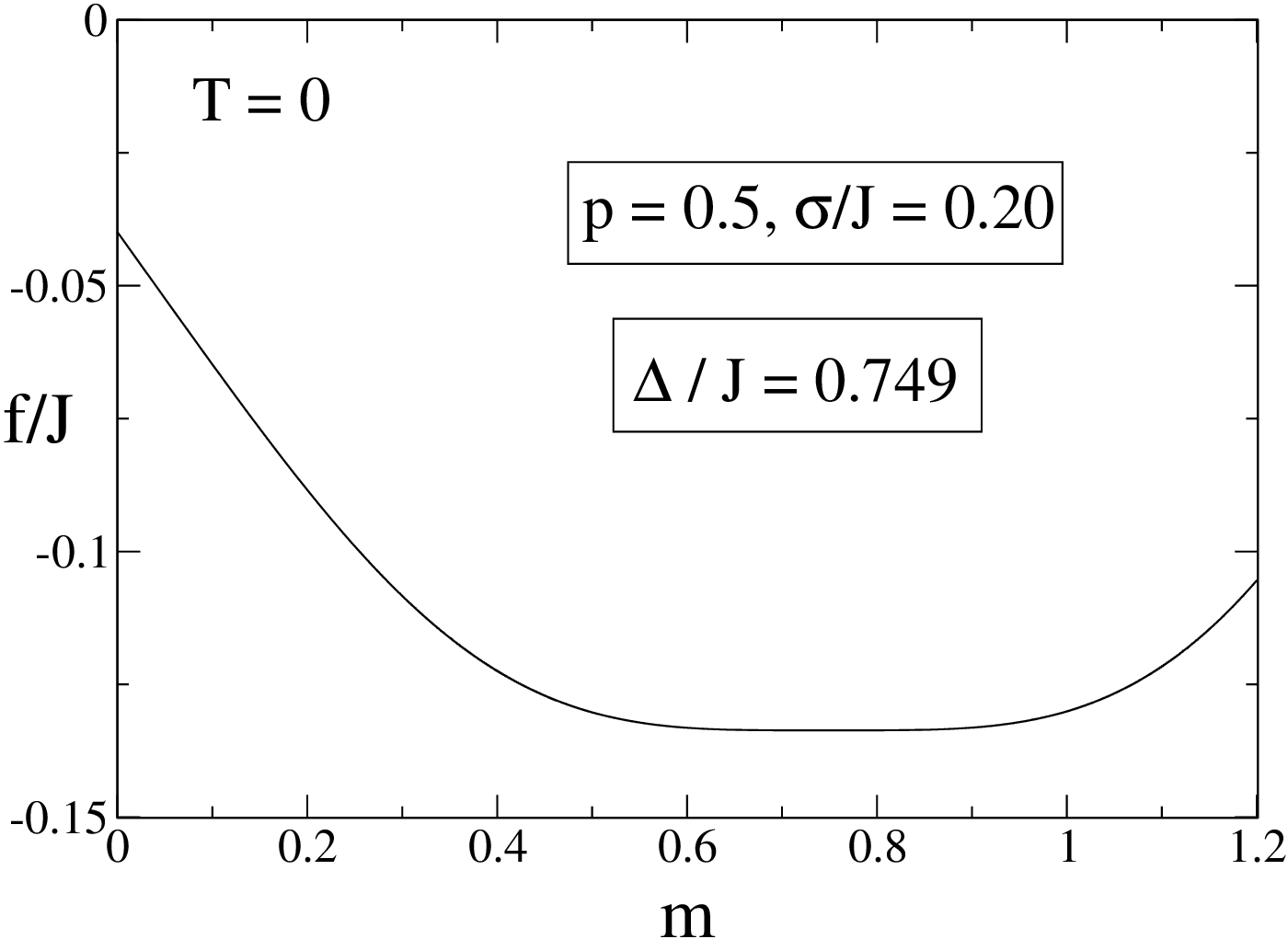}}
 
\caption{\footnotesize The Free energy (see Eq. (11)) versus the order parameter,  plotted for $p=0.5$, for three different values of $\sigma/J$. In (a) and (b) two ferromagnetic phases coexist, but in (c) we note that for $\sigma/J = 0.2$, there is already a crossover to pass from one ferromagnetic phase to the other. }

\label{imagenes6}
\end{figure}   
 
For completeness, Figure 7(a) shows  what Figure 6(c) illustrates by means of the free energy. There, it is shown
where the ordered critical point is located, that is, at $T=0$. In Figure 7(b), we see the line composed by the $(p,\sigma^{'}/J)$-points. This line separates phase diagrams containing two  and  one ferromagnetic phases.
Particularly, for $p=1$, $\sigma^{'}/J = (2\pi)^{-1/2}$, as  obtained in reference \cite{carneiro} and confirmed numerically by us.\\
We summarize the preceding analysis by showing, in Figure 8,  the regions of validity for the four qualitatively distinct phase diagrams. Note that along the horizontal axis ($\sigma/J=0$), regions II and III are separated by $p=0.8245$, and regions III and IV by $p=0.9258$. Along the vertical axis (at $p=1$), regions IV and III are separated by $\sigma/J=0.202$,  regions III and II by $\sigma/J=0.283$, then, regions II and I  by $\sigma/J=(2\pi)^{-1/2} \approx 0.3989$. Furthermore, the line separating topologies I and II is the same as in Figure 7(b). The frontier separating topologies II and III consists of points estimated by the analysis illustrated by  Figure 4. Finally, the line between topologies III and IV is made of fourth-order critical points, i.e., it is based upon the points in Figure 3.

\vskip \baselineskip  
 \begin{figure}[htp]
\centering
\vspace{0.7cm} 
\subfigure[][] {\includegraphics[width=5.6cm]{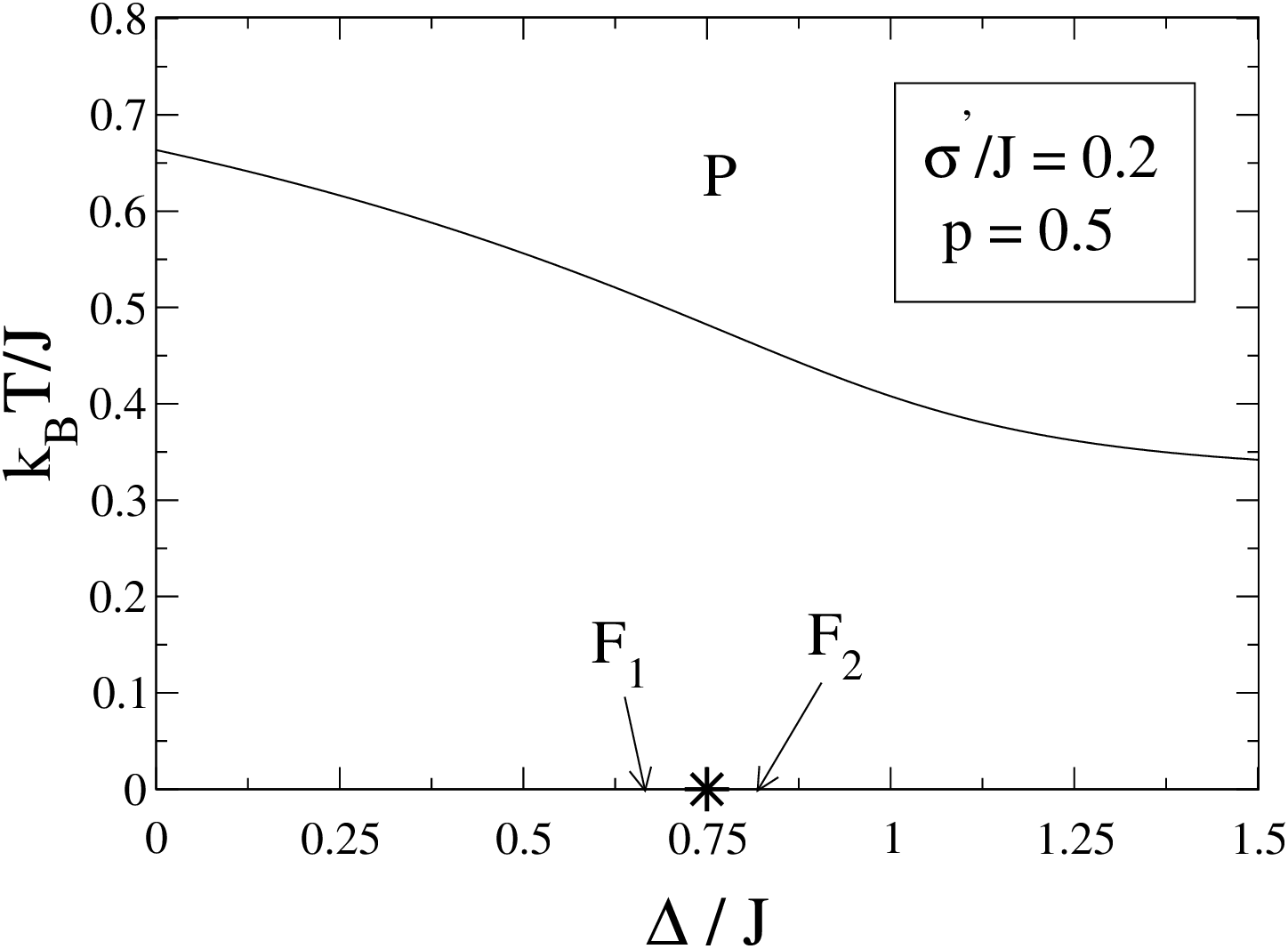}}
 \subfigure[][] {\includegraphics[width=5.6cm]{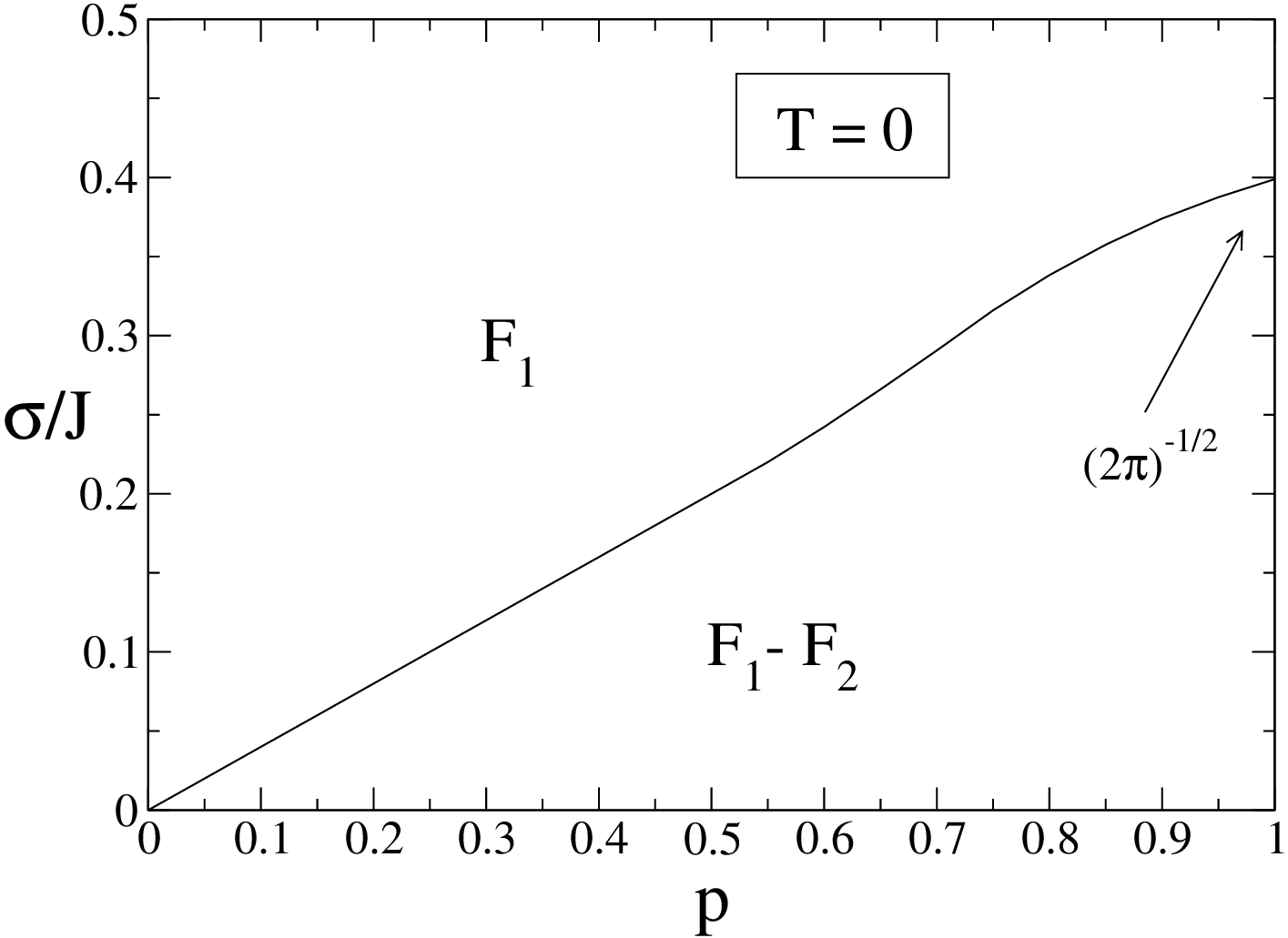}}
\caption{\footnotesize  In (a) is shown the phase diagram obtained for $p=0.5$, for the corresponding critical $\sigma^{'}/J$. Note that  the ordered-critical point (that appeared in Topology II) is now located at the horizontal axis. So, for $\sigma > \sigma^{'}$ there will be only one ferromagnetic order at low temperatures for $p=0.5$. In (b), the line separating topologies I and II. This line is made of points  numerically obtained by finding $\sigma^{'}/J$,  for each $p$.}
\label{imagenes9}
\end{figure}   
\vskip \baselineskip

\begin{figure}[tp]
\begin{center}
\centering
\vspace{0.7cm} 
\includegraphics[height=7cm]{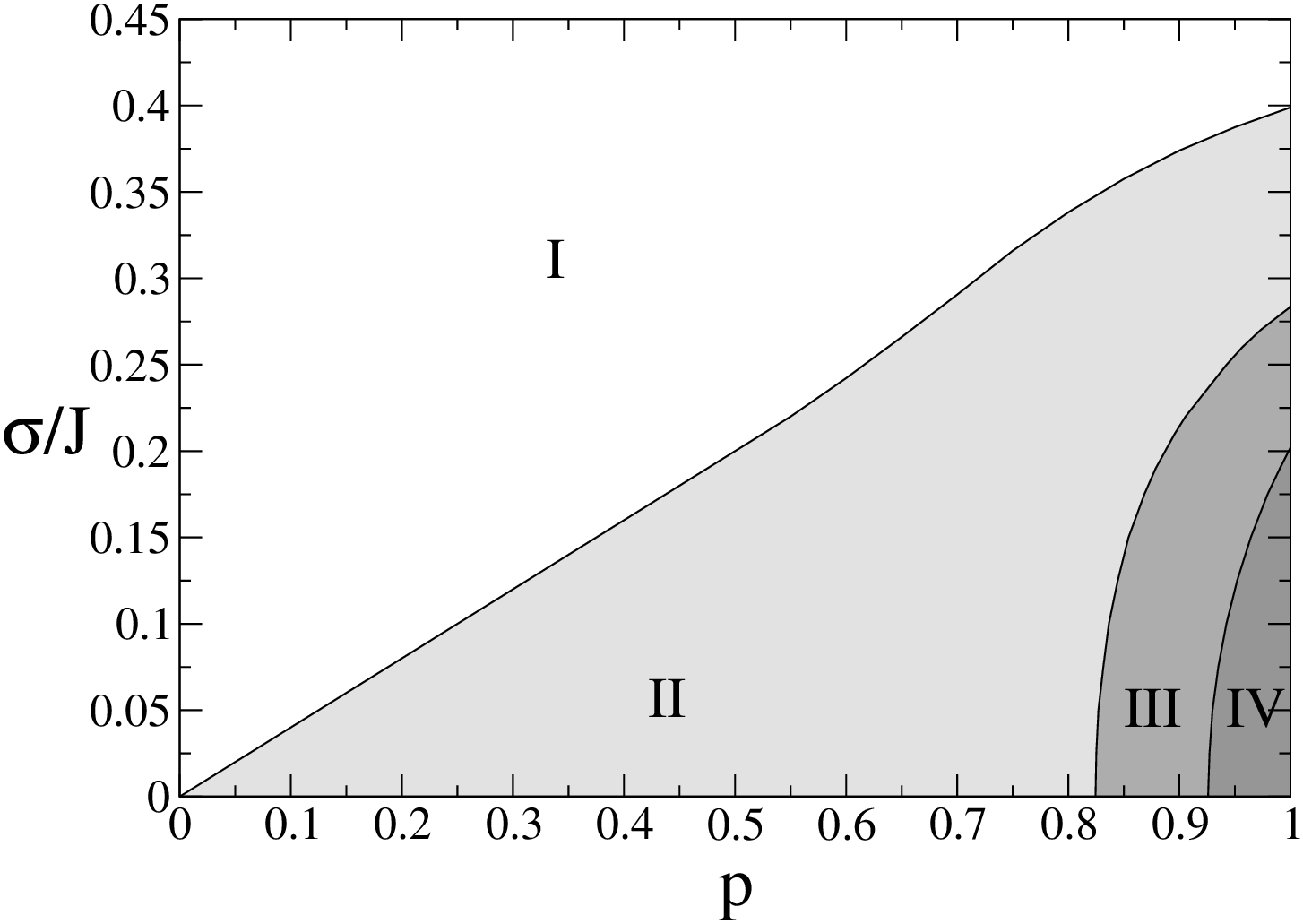}
\end{center}
\caption[]{\footnotesize Regions, in the plane $\sigma/J$ versus $p$, associated with the topologies for the present model (see also Figure 2). The horizontal  and the vertical axes represent the probability $p$, and the width $\sigma$, respectively (see Eq. (\ref{PDF})). The tricritical behavior belongs to the region IV. The  simplest topology belongs to region I, where  only one ferromagnetic phase  appears, whereas the rest topologies contain two ferromagnetic orders at low temperatures.}
\label{nxxx}
\end{figure}  
\vskip \baselineskip

\section{Conclusions}

We revisiting the study of the infinite-range-interaction spin-1 Blume-Capel Model with quenched randomness, by considering a more general
probability distribution function for the crystal field $\Delta_{i}$, which consists of two Gaussian distributions centered at   $\Delta_{i} = \Delta$ and $\Delta_{i} = 0$, with probabilities $p$ and $(1-p)$, respectively.
For $\sigma=0$, we recover the bimodal case studied in references \cite{benyoussef,salinas}, and for $p=1$, the
Gaussian case studied in  reference \cite{carneiro}. For $\sigma$-widths in $ 0 < \sigma < 0.202J $,  the system exhibits four distinct topologies according to the range in which  $p$ belongs. So, we designed them as Topology I,II,III, and IV,   in increasing order of $p$. Topology I contains one continuous critical line separating a ferromagnetic phase to the paramagnetic phase. In Topology II, one first-order critical line separating two ferromagnetic phases is added. This line terminates at an ordered critical point. The most complex criticality belongs to Topology III, where the first-order line now divides the continuous critical line by two critical end points. In Topology IV, the first-order line and the continuous line are met by a tricritical point. Accordingly, Topology I presents one ferromagnetic phase, whereas the rest ones show two distinct ferromagnetic orders at low temperatures. On the other hand,  the tricritical behavior  manifested in Topology IV emerges for $p>p^{*}$, where $p^{*}$ denotes
the probability for a given $\sigma/J$, where a fourth-order critical point is found. This point may be regarded as the last tricritical point vanishing  for  $\sigma/J > 0.202$, since $\sigma/J = 0.202$ leads to $p^{*} =1$. Consequently, the tricritical behavior is no longer  found  for any $p$.  Topology III disappears for $\sigma/J > 0.283$,  and Topology II is limited by $\sigma/J = 0.3989$, above which the first-order line separating the two ferromagnetic phases is suppressed for any $p$.  After that, for $\sigma/J > 0.3989$, only the simplest topology survives. \\\\
 Therefore,  we show through this model  how a complex magnetic criticality is reduced by the strength of the disorder (see also \cite{octavio,crokidakisa,crokidakisb}). Nevertheless, the critical dimensions for these types of phase diagrams is still an open problem to be solved.  
\vskip \baselineskip

{\large\bf Acknowledgments}

\vskip \baselineskip
\noindent
Financial support from
CNPq  (Brazilian agency) is acknowledged. 

\vskip 2\baselineskip

\end{document}